\documentstyle[12pt,epsf]{article}
\newcommand{\beq}{\begin{equation}}
\newcommand{\eeq}{\end{equation}}
\newcommand{\beqa}{\begin{eqnarray}}
\newcommand{\eeqa}{\end{eqnarray}}
\def\simgt{\rlap{\lower 4 pt \hbox{$\mathchar \sim$}} \raise 1pt \hbox {$>$}}
\def\simlt{\rlap{\lower 4 pt \hbox{$\mathchar \sim$}} \raise 1pt \hbox {$<$}}

\def\@citex[#1]#2{\if@filesw\immediate\write\@auxout{\string\citation{#2}}\fi
  \def\@citea{}\@cite{\@for\@citeb:=#2\do
    {\@citea\def\@citea{,\penalty\@m}\@ifundefined
       {b@\@citeb}{{\bf ?}\@warning
       {Citation `\@citeb' on page \thepage \space undefined}}%
\hbox{\csname b@\@citeb\endcsname}}}{#1}}
\def\citer{\@ifnextchar [{\@tempswatrue\@citexr}{\@tempswafalse\@citexr[]}}
\def\@citexr[#1]#2{\if@filesw\immediate\write\@auxout{\string\citation{#2}}\fi
  \def\@citea{}\@cite{\@for\@citeb:=#2\do
    {\@citea\def\@citea{--\penalty\@m}\@ifundefined
       {b@\@citeb}{{\bf ?}\@warning
       {Citation `\@citeb' on page \thepage \space undefined}}%
\hbox{\csname b@\@citeb\endcsname}}}{#1}}
\begin{document}

\begin{titlepage}
\begin{flushright}
        CERN-TH/98-120\\
        hep-ph/9804241 \hspace*{0.1cm}\\ 
        April 1998 \mbox{\hspace*{0.6cm}}\\
\end{flushright}
\vskip 1.8cm
\begin{center}
 \boldmath
{\Large\bf A quark mass definition adequate for\\[0.2cm]
threshold problems} 
\unboldmath
\vskip 2.2cm
{\sc M. Beneke}
\vskip .3cm
{\it Theory Division, CERN, CH-1211 Geneva 23, Switzerland}
\vskip 2.0cm
\end{center}

\begin{abstract}
\noindent 
Recent calculations of heavy quark cross sections near threshold 
at next-to-next-to-leading order have found second-order corrections 
as large as first-order ones. We analyse long-distance contributions 
to the heavy quark potential in momentum and coordinate space and 
demonstrate that long-distance contributions in momentum space are 
suppressed as $\Lambda_{\rm QCD}^2/\vec{q}^{\,2}$. We then show that 
the long-distance sensitivity of order $\Lambda_{\rm QCD}\vec{r}$ introduced 
by the Fourier transform to coordinate space cancels to all 
orders in perturbation theory with long-distance contributions to 
the heavy quark pole mass. This leads us to define a subtraction 
scheme -- the `potential subtraction scheme' -- in which large 
corrections to the heavy quark potential and the `potential-subtracted' 
quark mass are absent. We compute the two-loop relation of the 
potential-subtracted quark mass to the $\overline{\rm MS}$ quark mass. 
We anticipate that threshold calculations expressed in terms of 
the scheme introduced here exhibit improved convergence properties.  
\end{abstract}

\vfill

\end{titlepage}


{\bf \noindent 1. Motivation}\\[-0.2cm]

\noindent
Heavy quark production near threshold through virtual photons or 
$Z$ bosons is very sensitive to the quark mass and therefore may allow 
us to determine heavy quark masses precisely. Recently, the two-loop 
corrections to the colour Coulomb potential \cite{Pet97} and to 
the matching relation between the relativistic and non-relativistic 
vector current \cite{BSS98,CM98} were obtained. The two together provide 
the necessary input to compute heavy quark properties near threshold in 
next-to-next-to-leading order (NNLO). In this context `NNLO' means 
that all corrections to the Born cross section of order 
$(\alpha_s/v)^n\,\alpha_s^k$ for any $n$ and $k=0,1,2$ are taken 
into account, where $v$ is the small relative velocity 
of the two quarks in their centre-of-mass frame and $\alpha_s$ is the 
strong coupling. (Additional logarithms of $v$ are not written 
explicitly.)

NNLO calculations have now been completed for top-anti-top production 
near threshold \cite{HT98,MY98}, for bottomonium threshold sum rules 
\cite{PP98,Hoa98} and for quarkonium energy levels \cite{PY97}. 
In all three cases the NNLO correction is as large as the 
next-to-leading order (NLO) correction, which suggests that a perturbative 
treatment has already reached its limits. In case of $t\bar{t}$ 
production the NNLO correction shifts the location of the cross section 
peak position by about $1\,$GeV, which implies an uncertainty 
in $m_t$ of about $0.5\,$GeV, if the threshold cross section is used 
as a measurement of the top quark mass. This result is unexpected, in 
particular as the relevant physical scale is given by $C_F \alpha_s(m_t v) 
\,m_t/2\sim 15$-$20\,$GeV at which perturbation theory should work.

Following a different line of investigation, the Coulomb potential 
in momentum and in coordinate space is analysed in \cite{JKPST98,JPS98}. 
It is found (see also \cite{AL95}) that the effective couplings 
defined by the two versions of the potential are related by a 
rapidly divergent series, the origin of which is a long-distance 
contribution of relative order $\Lambda_{\rm QCD} r$. This leads 
to large numerical differences in different, but consistent at NNLO,  
implementations of the Coulomb potential in cross section calculations. 
The authors of \cite{JKPST98,JPS98} conclude, in agreement with the 
evidence from the NNLO $t\bar{t}$ and $b\bar{b}$ computations mentioned 
above, that there is a large and irreducible uncertainty that affects 
the threshold region. This would lead to rather gloomy prospects as to 
our ability to constrain the bottom and, eventually, the top quark 
mass.

In this paper we show that despite these evidences 
perturbation theory does not (yet) fail and that the actual 
uncertainties can be smaller than those indicated by 
\cite{HT98,MY98,Hoa98,JKPST98,JPS98}. 
Our main point is that, contrary 
to intuition, the notion of a quark pole mass, which has been 
implicit in the discussion above, is in fact inadequate for 
accurate calculations of heavy quark cross sections near 
threshold. The argument goes as follows:

We first analyse (Sect.~2) the heavy quark 
potential $\tilde{V}(\vec{q}\,)$ in momentum space and find that 
long-distance contributions have a relative suppression 
$(\Lambda_{\rm QCD}/\vec{q}\,)^2$. Hence the potential in 
momentum space is better behaved than the potential in coordinate 
space. Knowing that the long-distance contribution of relative 
order $\Lambda_{\rm QCD} r$ to the potential in coordinate space 
enters only through the Fourier transform, we can 
eliminate it by restricting the Fourier transform to 
$|\vec{q}\,|>\mu_f$ for some factorisation scale $\mu_f$ which 
satisfies
$\Lambda_{\rm QCD}<\mu_f<m v$. This defines a subtracted potential 
$V(r,\mu_f)$, from which large perturbative corrections  
are eliminated (Sect.~3). The Schr\"odinger equation takes its 
conventional form only if the pole quark mass definition is 
assumed. The Schr\"odinger equation formulated with the subtracted 
potential contains a residual mass term $\delta m(\mu_f)$. 
As a consequence the input parameter for threshold 
calculations in terms of the subtracted potential is not 
$m_{\rm pole}$ but $m_{\rm PS}(\mu_f)=m_{\rm pole}-\delta m(\mu_f)$. 
It is known that the pole mass also receives long-distance 
contributions of order $\Lambda_{\rm QCD}$ \cite{BB94,BSUV94} 
and it was noted already in \cite{BSUV94} that they are related 
to the Coulomb contribution to the self-energy. 
The crucial point is that the long distance 
sensitivity of order $\Lambda_{\rm QCD}$ 
in the coordinate space potential or, equivalently,  
$\delta m(\mu_f)$ cancels to all orders in perturbation theory with 
the leading long-distance sensitivity in the pole mass (Sect.~4). 
Hence the `potential-subtracted' mass 
$m_{\rm PS}(\mu_f)$ can be related to more conventional (long-distance 
insensitive) mass definitions by a well-behaved perturbative expansion. 
The two-loop relation to the $\overline{\rm MS}$ mass definition 
can be trivially obtained (Sect.~5). 

It follows from this argument that one can 
avoid large perturbative corrections as were found in the NNLO 
calculations mentioned above by formulating the threshold 
problem in terms of $m_{\rm PS}(\mu_f)$ and the subtracted potential 
$V(r,\mu_f)$ rather than the pole mass and the 
ordinary Coulomb potential. One 
can then determine $m_{\rm PS}(\mu_f)$ and relate it 
reliably to $m_{\overline{\rm MS}}$. The dependence on the 
factorisation scale $\mu_f$ cancels in this process.\\

{\bf \noindent 2. The potential in momentum space}\\[-0.2cm]

\noindent
The static potential in coordinate space, $V(r)$, is defined in terms of a 
Wilson loop $W_C(\vec{r},T)$ of spatial extension $\vec{r}$ and 
temporal extension $T$ with $T\to\infty$ \cite{Sus77,Fis77,ADM77}. 
In this limit $W_C(\vec{r},T)\sim \exp(-i T V(r))$. The potential 
in momentum space, $\tilde{V}(q)$, is the Fourier transform of 
$V(r)$. (We use $r=|\vec{r}\,|$ and $q=|\vec{q}\,|$.)  One can compute the 
potential directly in momentum space from the on-shell quark-anti-quark 
scattering amplitude (divided by $i$) at momentum transfer $\vec{q}$ 
in the limit of static quarks $m\to\infty$ and projected on the 
colour-singlet sector. In addition one has to 
change the sign of the $i\epsilon$-prescription of some of the 
anti-quark propagators in some diagrams (such as $D_1$ below), so 
that the integration over zero-components of loop momentum can always 
be done without encountering quark poles in the upper half plane. The 
quark poles amount to iterations of the potential, but do not give 
a contribution to the potential itself.\footnote{We 
remark that in the threshold expansion of Feynman 
integrals \cite{BS97} the static potential and its generalisation 
are generated by integrating out soft quarks and gluons with energy and 
three-momentum of order $m v$ together with potential gluons with energy 
of order $m v^2$ and three-momentum of order $m v$.} 

The potential is infrared (IR) finite\footnote{We are aware of the 
analysis of \cite{ADM77}, which can be interpreted as a statement to 
the contrary. In our opinion, the interpretation of the divergences 
discussed in \cite{ADM77} deserves further consideration.} and 
ultraviolet (UV) finite 
after renormalisation of the coupling. In this Section we ask how 
sensitive the Feynman integrals 
that contribute to the momentum space potential are to the IR regions 
of loop momentum integrations. The reason is 
that these regions give rise to large corrections in higher orders 
in perturbation theory through IR renormalons (see \cite{BB94,BSUV94} 
for references). Note that we are not concerned with the 
long-distance behaviour of the potential at $q\sim \Lambda_{\rm QCD}$, 
but with the leading power corrections of form $(\Lambda_{\rm QCD}/q)^k$,  
which correct the perturbative Coulomb potential when $q$ is still 
large compared to $\Lambda_{\rm QCD}$.

\begin{figure}[t]
   \vspace{-4.5cm}
   \epsfysize=30cm
   \epsfxsize=20cm
   \centerline{\epsffile{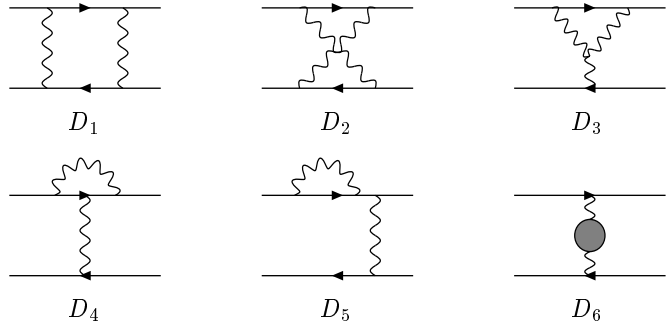}}
   \vspace*{-20.7cm}
\caption[dummy]{\label{fig1}\small 
One-loop corrections to the heavy quark potential.}
\end{figure}
Consider first the one-loop corrections to the potential (see 
Fig.~\ref{fig1}) in dimensional regularisation. In Feynman gauge $D_3$ 
is zero. The colour $C_F^2$-term cancels in the sum $D_1+D_2$. $D_{1,2}$ are  
logarithmically IR divergent. This IR divergence is cancelled by the 
(logarithmic) scaleless integrals $D_{4,5}$, whose only effect is to 
convert the 
IR singularity in $D_{1,2}$ into a UV singularity, which can be absorbed 
into a renormalisation of $\alpha_s$. Hence, we are left with the 
IR finite parts of (the $C_A C_F$-part of) $D_2$ and $D_6$, which 
yield the well-known 
one-loop correction to the potential \cite{Fis77,ADM77}. $D_6$ is 
given in terms of the gluon self-energy at off-shell 
external momentum $\vec{q}$. For small loop momentum $k$, the 
integral scales as 
\begin{equation}
D_6\sim \frac{1}{q^4}\int d^4 k \,\frac{q^2}{k^2 q^2}
\end{equation}
and gives rise to a contribution of order $\Lambda_{\rm QCD}^2/q^2$ 
from the region $k\sim \Lambda_{\rm QCD}$ relative to the tree diagram 
which scales as $1/q^2$. The integral relevant for $D_2$ is 
\begin{equation}
\label{d2int}
\int d^4 k \,\frac{1}{k^2 (k+q)^2 (v\cdot k)^2},
\end{equation}
where $v=(1,\vec{0}\,)$ and $v\cdot q=0$. To find the leading 
infrared contribution, which is left over after the 
IR divergence is cancelled as described above, we expand the integrand 
in $k$ around $k=0$ and around $k+q=0$. The integrals in each term 
of the expansion depend only on the 
vector $v$. Hence, in a regularisation scheme that preserves Lorentz 
invariance all odd terms vanish because $v\cdot q=0$. The 
long-distance contribution is again of relative order 
$\Lambda_{\rm QCD}^2/q^2$. (Note that we are only concerned with 
IR contributions that are connected with the large-order behaviour 
of perturbative expansions.)

This argument generalises to an arbitrary diagram. Because $v\cdot q=0$ 
and because there is no other kinematic invariant linear in $q$, 
it follows from Lorentz invariance that the leading power correction 
to the potential in momentum space cannot be $\Lambda_{\rm QCD}/q$, 
but has to be quadratic:
\begin{equation}
\label{potmom1}
\tilde{V}(q) = -\frac{4\pi C_F\alpha_s(q)}{q^2}\left(
1+\frac{\alpha_s(q)}{4\pi}\left[\frac{31}{3}-\frac{10 n_f}{9}\right]
+\ldots + \,{\rm const}\times\frac{\Lambda_{\rm QCD}^2}{q^2} + 
\ldots\right).
\end{equation}
The implication is that the expansion of the potential in the 
coupling\footnote{In this paper $\alpha_s(\mu)$ denotes the coupling 
defined in the $\overline{\rm MS}$ scheme.} $\alpha_s(q)$
diverges as $\sum_n r_n\alpha_s(q)^{n+1} \sim \sum_n (-a\beta_0)^n n! 
n^b\alpha_s(q)^{n+1}$ 
with $a=1$ and $\beta_0=-(11-2 n_f/3)/(4 \pi)$ the first coefficient 
of the $\beta$-function. A linear IR power correction would have led 
to $a=2$ and, hence, a more rapidly divergent perturbative relation. 
The parameter $b$ remains undetermined by the above analysis, but 
does not influence the power behaviour.\\

{\bf \noindent 3. The potential in coordinate space}\\[-0.2cm]

\noindent
Consider now the potential in coordinate space, defined by the 
Fourier transform
\begin{equation}
\label{fourier}
V(r) = \int\!\frac{d^3\vec{q}}{(2\pi)^3}\,\,e^{i\vec{q}\cdot\vec{r}}\,
\tilde V(q).
\end{equation}
Note that in the Wilson loop definition of the potential in 
coordinate space there is a (divergent) constant related to the 
self-energy of the static sources. This $r$-independent term is usually 
discarded when one refers to the static potential, and it is 
also discarded, when $V(r)$ is defined in terms of the Fourier 
integral above. In what follows the interpretation of this constant 
plays an important role.

To see that the potential in coordinate space is more sensitive 
to long distances than the potential in momentum space, it is enough 
to take the tree approximation to $\tilde{V}(q)$ and to 
calculate
\begin{equation}
\int\limits_{|\vec{q}\,|<\Lambda_{QCD}} 
\!\!\!\!\frac{d^3\vec{q}}{(2\pi)^3}\,\,e^{i\vec{q}\cdot\vec{r}}\,
\left(-\frac{4\pi C_F\alpha_s}{q^2}\right) = 
-\frac{2 C_F\alpha_s}{\pi}\,\Lambda_{\rm QCD} + 
O(\Lambda_{\rm QCD}^3 r^2).
\end{equation}
It follows that the leading power correction is linear in 
$\Lambda_{\rm QCD} r$:
\begin{equation}
V(r) = -\frac{C_F\alpha_s(e^{-\gamma_E}/r)}{r}\left(
1+\frac{\alpha_s(e^{-\gamma_E}/r)}{4\pi}\left[\frac{31}{3}-
\frac{10 n_f}{9}\right]
+\ldots + \,{\rm const}\times\Lambda_{\rm QCD} r + 
\ldots\right).
\end{equation}
($\gamma_E$ is Euler's constant.) 
The implication is that the expansion of the coordinate space 
potential in $\alpha_s(e^{-\gamma_E} r)$ diverges as 
$\sum_n r_n\alpha_s(e^{-\gamma_E} r)^{n+1} \sim \sum_n (-a\beta_0)^n n! 
n^b\alpha_s(e^{-\gamma_E}/r)^{n+1}$ 
with $a=2$, much faster than the expansion of the potential in 
momentum space. Note that in absolute terms the IR contribution 
is a constant of order $\Lambda_{\rm QCD}$.

The rapidly divergent behaviour of the coordinate space potential 
has been noted in \cite{AL95} and is discussed in detail in 
\cite{JKPST98,JPS98}. What we add here is the observation that 
the linear power correction and, by implication, the rapid divergence 
originates {\em only} from the Fourier transform to coordinate space 
and is not present in the potential in momentum space. Knowing this 
we can subtract the leading long-distance contribution and 
the leading divergent behaviour completely by restricting the 
Fourier integral to $|\vec{q}\,|>\mu_f$ with $\mu_f$ a factorisation 
scale which we make more precise later. The result will be 
called the `subtracted potential' $V(r,\mu_f)$. The subtraction terms 
can be evaluated order by order in $\alpha_s$ given $\tilde{V}(q)$ 
to that order. The relevant calculation will be done in Sect.~5. 
To be precise, we write
\begin{equation}
V(r,\mu_f) = V(r)+2\delta m(\mu_f),
\end{equation}
where 
\begin{equation}
\label{deltam}
\delta m(\mu_f) = -\frac{1}{2}\int\limits_{|\vec{q}\,|<\mu_f} 
\!\!\!\frac{d^3\vec{q}}{(2\pi)^3}\,\tilde{V}(q).
\end{equation}
To subtract the leading long-distance contribution of order 
$\Lambda_{\rm QCD}$, it is legitimate to replace 
$e^{i\vec{q}\cdot\vec{r}}$ by 1 in the Fourier transform and 
we use this for the definition of the subtraction term. 
We now define the `potential subtracted' 
(PS) quark mass parameter as
\begin{equation}
m_{\rm PS}(\mu_f) = m_{\rm pole}-\delta m(\mu_f).
\end{equation}
In terms of the subtracted potential and the PS mass the Green function 
is determined from the Schr\"odinger-type equation
\vspace*{0.1cm}
\begin{equation}
\left[-\frac{\nabla^2}{m_{\rm PS}(\mu_f)} + V(r,\mu_f) 
- E\right] G_C(\vec{r},E) = \delta^{(3)}(\vec{r}\,), 
\vspace*{0.1cm}
\end{equation}
and it is important that the non-relativistic energy is defined 
as $E=\sqrt{s}-2 m_{\rm PS}(\mu_f)$, with $s$ the centre-of-mass energy, 
as compared to the usual definition $\sqrt{s}-2 m_{\rm pole}$. 
The equation above is the conventional 
Schr\"odinger equation, but with a `residual' mass term.\footnote{We  
have replaced $m_{\rm pole}$ by $m_{\rm PS}(\mu_f)$ also in the 
kinetic energy term in the Schr\"odinger equation. The difference 
to using the pole mass is $\delta m/m\, \nabla^2$, i.e. of higher order 
in $1/m$. Since other terms of order $1/m$ are neglected in the 
Schr\"odinger equation, the replacement of $m_{\rm pole}$ by 
$m_{\rm PS}(\mu_f)$ is justified. If one chooses the PS mass 
as the mass definition and derives the Schr\"odinger equation 
from Feynman diagrams, the kinetic energy term is divided by 
the PS mass by construction. An additional finite renormalization 
of the kinetic energy term can be neglected in the present context.}
The subtracted potential $V(r,\mu_f)$, which contains the 
residual mass, does not suffer from 
large loop corrections associated with the leading asymptotic 
behaviour of its perturbative expansion.

Despite this fact we have not yet gained anything, because the 
large loop corrections have only been hidden in the contribution 
$\delta m(\mu_f)$ to $m_{\rm PS}(\mu_f)$. The crucial point is this:  
When $m_{\rm pole}$ is expressed in terms of a 
`short-distance' mass parameter such as the $\overline{\rm MS}$ 
mass $m_{\overline{\rm MS}}$ through a perturbative series, 
this perturbative 
series also has large loop corrections \cite{BB94,BSUV94}. The 
large perturbative corrections absorbed into 
$\delta m(\mu_f)$ cancel exactly with large perturbative corrections 
to the pole mass. The argument is given in the following Section.
Hence one can first determine the PS mass from the threshold cross 
section without encountering large corrections, because of 
the subtraction in the potential. One can then relate 
the PS mass to the $\overline{\rm MS}$ mass, again without 
encountering large corrections related to the asymptotic behaviour 
of perturbative expansions. In this way, one can, 
in principle, determine the 
$\overline{\rm MS}$ mass from threshold cross sections to better 
accuracy than the pole mass, the use of which is implied by 
the unsubtracted potential. Conversely, one can begin with 
$m_{\overline{\rm MS}}$, compute the PS mass and predict the threshold 
behaviour. The perturbative relation between the PS mass and 
the $\overline{\rm MS}$ mass is given in Sect.~5. 

One may ask why we do not suggest to avoid using the coordinate space 
potential and to work with the momentum space potential directly. 
The reason is that the Schr\"odinger equation for the Coulomb Green 
function $\tilde{G}_C(\vec{p},E)$ in momentum space contains the 
integration
\begin{equation}
\int\frac{d^3 k}{(2\pi)^3}\,\tilde{V}(\vec{k})\,
\tilde{G}_C(\vec{p}-\vec{k},E), 
\end{equation}
which contains exactly the same leading long-distance sensitivity 
as the Fourier integral (\ref{fourier}), because 
$e^{i\vec{q}\cdot \vec{r}}$ may 
be replaced by 1 as far as the leading power in $\Lambda_{\rm QCD} r$ 
is concerned, cf.~(\ref{deltam}). Hence the problem one 
encounters with the unsubtracted 
coordinate space potential enters in momentum space 
when one solves the Schr\"odinger equation.

How large can $\mu_f$ be? We shall see below that the expansion of 
$\delta m(\mu_f)$ is naturally expressed in terms of $\alpha_s(\mu_f)$. 
Perturbativity hence requires $\mu_f >\Lambda_{\rm QCD}$. The scale 
relevant to the potential is $1/r\sim m v$. Since the subtraction 
should affect the potential only at distances larger than the physical 
scale of the process described by the potential we require also 
$\mu_f< m v$. There is another way to arrive at this constraint. 
If $v=\sqrt{1-4 m_{\rm pole}^2/s}$ is expanded in terms of 
$m_{\rm PS}(\mu)$, one generates singular terms (as $v\to 0$) of 
order $(\delta m(\mu_f)/(m v^2))^k$. These terms are small if 
$\delta m(\mu_f)$ is small compared to scale $m v^2$ of binding energies 
of a Coulomb 
system. Counting $\alpha_s\sim v$ and using 
$\delta m(\mu_f)\sim \mu_f\alpha_s$, 
one arrives again at $\mu_f < m v$.\\

{\bf \noindent 4. Cancellation of long-distance 
contributions with the pole mass}\\[-0.2cm]

\noindent
Expressing the pole mass in terms of the $\overline{\rm MS}$ mass 
and $\delta m(\mu_f)$ as a series in $\alpha_s$, we can write 
\begin{equation}
\label{psms}
m_{\rm PS}(\mu_f) = m_{\rm pole}-\delta m(\mu_f) = 
M\left[1+\sum_{n=0} r_n 
\alpha_s^{n+1}\right] - \mu_f \sum_{n=0} s_n \alpha_s^{n+1},
\end{equation}
where $M=m_{\overline{\rm MS}}(m_{\overline{\rm MS}})$.
Both series diverge as 
$r_n,s_n \sim (-a\beta_0)^n n! n^b$ with $a=2$. We now show 
that this behaviour cancels in the difference in (\ref{psms}).
Because this divergence arises from long-distance sensitivity of 
order $\Lambda_{\rm QCD}$ in the Feynman diagrams that contribute 
to the two series, it is enough to show that the corresponding 
linear\footnote{A loop integral that behaves as 
$\int d^4k/k^4$ for small $k$ is called logarithmically IR sensitive, 
an integral that behaves as $\int d^4k/k^3$ linearly IR 
sensitive etc..} IR behaviour of the Feynman integrands cancels to all orders 
in perturbation theory in the difference. The remaining long-distance 
contributions to the difference are of order $\Lambda_{\rm QCD}^2/M$ 
and the corresponding divergent behaviour has only $a=1$. This 
establishes that the PS mass can be reliably related to the 
$\overline{\rm MS}$ mass. 

Since we are only concerned with infrared behaviour we may work with 
unrenormalised quantities which differ from $\overline{\rm MS}$ 
renormalised quantities only by pure UV subtractions. 
Radiative corrections to the pole mass are given by the 
self-energy, evaluated at $p^2=m_{\rm pole}^2$: 
$\Delta m\equiv m_{\rm pole}-m_0 = \Sigma_{|\not p=m_{\rm pole}}^0$, 
where $m_0$ is the bare mass and $\Sigma^0$ the unrenormalised 
self-energy.\footnote{The self-energy is given by the one-particle 
irreducible two-point diagrams divided by $(-i)$.} Consider 
the cancellation at the 1-loop order. As long 
as we are interested only in the leading behaviour at small loop 
momentum, we can approximate the one-loop contribution 
to $\Sigma^0$ by 
\begin{equation}
\label{ints}
-iC_F g_s^2\int\!\frac{d^4 q}{(2\pi)^4}
\frac{\gamma^\mu(\not\!p+\!\not\!q+m_0)\,\gamma_\mu}{((p+q)^2-m_0^2)\,q^2} 
\Big|_{p^2=m_0^2} \longrightarrow 
-iC_F g_s^2\int\!\frac{d^4 q}{(2\pi)^4} 
\frac{1}{(v\cdot q+i\epsilon) \,(q^2+i\epsilon)},
\end{equation}
setting $p=m_0 v$. At this order one need not distinguish between 
$m_{\rm pole}$ and $m_0$ in the integrand. 
Taking the integration over $q_0$ in the upper 
half plane, we obtain
\begin{equation}
-\frac{1}{2} \int\frac{d^3 \vec{q}}{(2\pi)^3} \frac{4\pi C_F \alpha_s}
{-\vec{q}^{\,2}},
\end{equation}
which is exactly the leading-order contribution to $\delta m(\mu_f)$, 
cf.~(\ref{deltam}), for small $\vec{q}$. 
Thus the Feynman integrands of the integrals 
contributing to $M r_0$ and $\mu_f s_0$ cancel each other 
in the infrared region of small $q$. Note that the leading infrared 
behaviour can be obtained by replacing
\begin{equation}
\frac{1}{v\cdot q+i\epsilon} \longrightarrow -i\pi\,\delta(v\cdot q)
\end{equation}
in (\ref{ints}). In other words, the relevant IR behaviour is 
obtained from setting $q_0=0$ first and then from the small-$\vec{q}$ 
behaviour of the remaining three-dimensional integral.

The denominator of an on-shell heavy quark propagator with 
momentum $m v+l$ is $v\cdot l+l^2/(2 m)$. To demonstrate the 
IR cancellation in higher-loop order, we consider first the 
static approximation, in which the denominator is simplified to 
$v\cdot l$ and the gluon coupling 
$\gamma_\mu$ to heavy quarks by $v_\mu$. Hence the Feynman rules 
reduce to those implicit in the definition of the potential. 
The static approximation implies 
that we consider all loop momenta small compared 
to $m$ and take the first term in an expansion in $1/m$. We show that 
the leading IR contributions of order $\Lambda_{\rm QCD}$ 
to the pole mass in this approximation cancel exactly with those 
to the coordinate space potential. At the end of this Section, 
we address the question of what happens, when one includes 
further terms in the expansion of the heavy quark 
propagator in $l$ and the region $l\sim m$, in which 
the propagator cannot be expanded. 

\begin{figure}[t]
   \vspace{-5.7cm}
   \epsfysize=30cm
   \epsfxsize=20cm
   \centerline{\epsffile{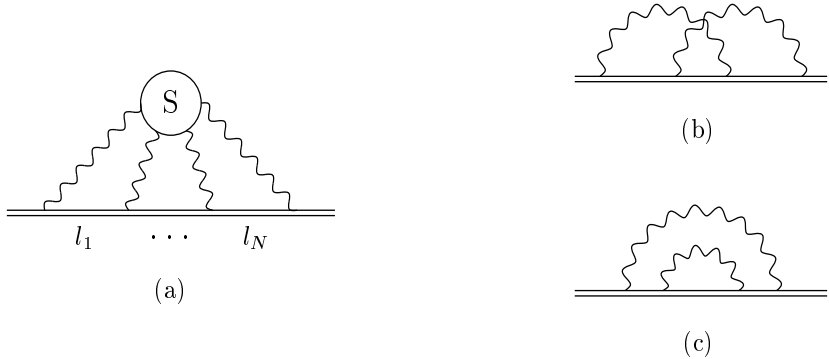}}
   \vspace*{-19.2cm}
\caption[dummy]{\label{fig2}\small 
(a) Structure of an arbitrary self-energy diagram. (b,c) Some 2-loop 
examples. The double line denotes the static approximation.}
\end{figure}

A general self-energy diagram in the static 
approximation can be written as (see 
Figure~\ref{fig2}a) 
\begin{equation}
\label{general}
\int\!\prod_{m=1}^L\frac{d^4 k_m}{(2\pi)^4}\,S(l_i)\,
\prod_{i=1}^N \frac{1}{v\cdot l_i+i\epsilon}
\end{equation}
where the line momenta $l_i$ are linear combinations of the loop 
momenta $k_m$ and $S(l_i)$ contains no heavy quark propagators. 
Consider any one-particle irreducible (1PI) subgraph 
contained entirely in $S$ at fixed, non-zero, external momentum. 
Such subgraphs are IR finite and at most quadratically IR 
sensitive. It follows that any subgraph that can give rise to 
linearly IR sensitive integrals must contain at least one 
heavy quark propagator with $v\cdot l_i\to 0$. 
Let us call a self-energy diagram 
(ir-)reducible if it contains (does not contain) a self-energy 
subgraph.

Consider irreducible diagrams such as the diagram depicted in 
Figure~\ref{fig2}b first. Irreducible 
diagrams are IR finite on mass-shell and none of the heavy quark line 
momenta coincide. The leading 
IR behaviour is obtained by letting one of the $v\cdot l_i=l_i^0$ 
go to zero and by neglecting $l_i^0$ in the other propagators. 
This can be summarised as the substitution
\begin{equation}
\label{irapprox}
\prod_{i=1}^N \frac{1}{v\cdot l_i+i\epsilon}
\,\longrightarrow\,
\sum_{j=1}^N \,(-i\pi)\,\delta(v\cdot l_j)\!\prod_{i=1,\,i\,\not\!=j}^N 
\frac{1}{v\cdot l_i+i\epsilon}.
\end{equation}
In a term with $\delta(v\cdot l_j)$ change one loop integration 
variable to $q=l_j(k_m)$. The delta-function kills the $q_0$ 
integral and sets $q_0=0$ in the integrand. One can interpret 
(\ref{irapprox}) as cutting the diagram at any heavy quark 
propagator. This gives rise to $N$ four-point diagrams,  
to be integrated with $d^3\vec{q}/(2\pi)^3$, and it is easy to 
see that the result matches exactly with contributions 
to $\delta m(\mu_f)$. For example, the diagram of Figure~\ref{fig2}b 
cancels with the contribution to $\delta m(\mu_f)$ from 
$D_2+2\cdot D_4$ in Figure~\ref{fig1}. The signs and $i$'s work 
out correctly and the factor $1/2$ in (\ref{deltam}) comes 
from the fact that we have $\pi$ in (\ref{irapprox}) but 
one factor $1/(2 \pi)$ from the four-dimensional integration 
measure.

One may be concerned about the fact that after setting 
$q_0=0$, the integrals become IR divergent and that hence 
the leading-order IR approximation (\ref{irapprox}) may  
not be sufficient for linearly IR sensitive contributions. 
The IR divergences are just the IR divergences in individual 
contributions to the potential mentioned in 
Sect.~2, which cancel in combinations such as $D_2+2\cdot D_4$. 
Moreover, the next term in the small-loop momentum expansion 
of the integrand is quadratically IR sensitive as 
shown in Sect.~2, and this is enough to guarantee that 
(\ref{irapprox}) is legitimate. 

One may also be concerned about the fact that application of 
(\ref{irapprox}) does not lead to $D_2$ literally, but to an 
integrand which differs from (\ref{d2int}) in that the two 
factors of $v\cdot k$ have different $i\epsilon$-prescriptions. 
However, in Sect. 2 we have shown that linearly IR sensitive 
contributions originate only from the IR behaviour of the 
Fourier integral. Hence we should consider $q$ small at fixed 
$k$ or both $q$ and $k$ small simultaneously (but not small 
$k$ at fixed $q$), in which case the potential pinch 
singularity is not a problem.

The situation is more complicated for reducible diagrams such 
as the one in Figure~\ref{fig2}c. For reducible diagrams some 
of the heavy quark line momenta coincide and cutting such a 
quark line in the sense discussed above leads to one-particle 
reducible contributions to the potential, i.~e. to lower-order 
contributions to the potential multiplied by on-shell 
renormalisation of the external legs. ($D_5$ in Figure~\ref{fig1} 
is an example.) Moreover, reducible diagrams 
are IR divergent when evaluated on-shell, 
while $\Delta M$ is IR finite. This is related 
to the fact that $\Delta M$ is not given by 
$\Sigma_{|\not p=m_0}^0$ but by $\Sigma_{|\not p=m_{\rm pole}}^0$.
To make the IR finiteness explicit, the contributions 
from reducible diagrams to $\Delta M$ should be 
combined with contributions at the same order in perturbation theory 
that arise from expanding $\Sigma_{|\not p=m_{\rm pole}}^0$ in 
$\Delta M$: 
\begin{eqnarray}
\label{iteration}
\Sigma_{|\not p=m_{\rm pole}}^0 &=& \Sigma_{|\not p=m_0}^0 + 
\Sigma_{|\not p=m_0}^0\,\frac{\partial\Sigma^0}{\partial\!\!\not\!p}
|_{\not p=m_0} 
+\Sigma_{|\not p=m_0}^0\left(\frac{\partial\Sigma^0}{\partial\!\!\not\!p}
|_{\not p=m_0}\right)^2
\nonumber\\
&&\hspace*{-1cm} +\,\frac{1}{2} \left(\Sigma_{|\not p=m_0}^0\right)^2 
\,\frac{\partial^2\Sigma^0}{\partial\!\!\not\!p^2}
|_{\not p=m_0}+ \ldots.
\end{eqnarray}
This expansion reproduces precisely the combinatorial structure 
of self-energy subgraphs in reducible diagrams. Combining the 
various terms on the level of integrands, the resulting 
integral is IR finite. Moreover, after this cancellation 
all heavy quark propagators have different momenta and one 
can again use (\ref{irapprox}). The `cut' diagrams then cancel 
again with diagrams to the potential.

Let us illustrate this for the diagram of Figure~\ref{fig2}c. 
According to (\ref{iteration}) we combine the diagram with 
the product of 1-loop contributions to the second 
term\footnote{Further terms contribute only at 3-loop order and beyond.} 
on the right hand side of (\ref{iteration}). 
This gives the following contribution 
to $\Delta M$:
\begin{eqnarray}
&&-g_s^4\int\!\frac{d^4k_1}{(2\pi)^4}\frac{d^4k_2}{(2\pi)^4} 
\left[\frac{1}{k_1^2 \,k_2^2\,(v\cdot k_1)^2\,v\cdot (k_1+k_2)}
-\frac{1}{k_1^2 \,k_2^2\,(v\cdot k_1)^2\,v\cdot k_2}\right]
\nonumber\\
&& = g_s^4 \int\!\frac{d^4k_1}{(2\pi)^4}\frac{d^4k_2}{(2\pi)^4} 
\frac{1}{k_1^2 \,k_2^2\,v\cdot k_1\,v\cdot (k_1+k_2)\,v\cdot k_2}
\\
&& \longrightarrow \,
-\frac{1}{2}\int\!\frac{d^3 \vec{q}}{(2\pi)^3}\,(-i)g_s^4 
\int\!\frac{d^4k}{(2\pi)^4}\left[\frac{-2}{k^2\,q^2\,(v\cdot k)^2} + 
\frac{1}{k^2 (k+q)^2\,(v\cdot k)^2}\right]
\nonumber\\
&& = -\frac{1}{2}\int\!\frac{d^3 \vec{q}}{(2\pi)^3}\,\frac{1}{i}
\left[\,4\cdot\frac{1}{2}\cdot D_5+ D_1\right].
\nonumber
\end{eqnarray}
To arrive at the last two lines we have used (\ref{irapprox}). 
Since $1/2 \cdot D_5$ is the on-shell wave function renormalisation 
for a single external quark leg times the leading-order potential, 
we obtain the desired cancellation with IR contributions to the 
potential. The example and the structure of (\ref{iteration}) 
make it transparent how the IR cancellation for reducible 
diagrams extends to all orders.

Let us return to the validity of the static approximation. Consider 
first the contributions to the pole mass from the region of loop 
momentum where all momenta are small compared to $m$. All heavy quark 
propagators can be expanded about the static limit. The corrections 
to the static approximation are suppressed by at least one power 
of loop momentum divided by $m$. This implies a suppression of 
long-distance sensitivity by a factor of $\Lambda_{\rm QCD}/m$ relative 
to the leading term. Since the leading term is already linear 
in $\Lambda_{\rm QCD}$, we conclude that if the heavy quark line 
momentum is small, it is sufficient to keep only the leading 
term in the expansion of the propagator. Consider now the contributions 
from the region of loop momentum where some loop momenta are 
of order $m$ and others (at least one) are small compared to 
$m$. The hard subgraphs with loop momentum of order $m$ reduce 
to local interactions of form $(l_i/m)^k$ with respect to the small 
loop momenta $l_i\sim \Lambda_{\rm QCD}$. One obtains contributions 
suppressed by powers of $\Lambda_{\rm QCD}/m$ unless $k<1$. Hence 
we need to consider only hard self-energy and vertex subgraphs. 
The effect of these subgraphs is to renormalise the coefficients 
of the $\psi^\dagger \psi A_0$ and $\psi^\dagger\partial_0\psi$ interaction 
terms in the non-relativistic effective Lagrangian, from which 
the potential is derived. In the standard normalisation of 
the non-relativistic Lagrangian the coefficients of these 
operators are 1 to all orders in 
perturbation theory. It follows that the hard subgraphs 
have no effect on the IR cancellation, or, in other words, 
they are implicitly taken into account through the 
coefficient functions of the interaction terms in the 
non-relativistic Lagrangian which enter the calculation of the 
Coulomb potential.\\
 
{\bf \noindent 5. Relation to the $\overline{\rm MS}$ mass 
definition}\\[-0.2cm]

\noindent
It is straightforward to work out the mass subtraction 
$\delta m(\mu_f)$ from known results on the potential in momentum 
space:
\begin{equation}
\tilde{V}(\vec{q}\,) = -\frac{4\pi C_F\alpha_s(\vec{q}\,)}{\vec{q}^{\,2}} 
\left[1+a_1 \,\frac{\alpha_s(\vec{q}\,)}{4\pi}+a_2\left(
\frac{\alpha_s(\vec{q}\,)}{4\pi}\right)^2\right]
\end{equation}
with $a_1$ as in (\ref{potmom1}) and $a_2=634.402-66.3542 n_f+1.246 n_f^2$ 
\cite{Pet97}. From the definition (\ref{deltam}) one obtains 
\begin{eqnarray}
\delta m(\mu_f) &=& \frac{C_F\alpha_s(\mu)}{\pi}\,\mu_f\Bigg[
1+\frac{\alpha_s(\mu)}{4\pi}\left(a_1-b_0\left(\ln\frac{\mu_f^2}{\mu^2} 
-2\right)\right)
\\
&&\hspace*{-1.5cm}+\left(\frac{\alpha_s(\mu)}{4\pi}\right)^2 
\Bigg(a_2-\left\{2 a_1 b_0+b_1\right\}\left(\ln\frac{\mu_f^2}{\mu^2} 
-2\right)
+b_0^2\left(\ln^2\frac{\mu_f^2}{\mu^2}-4 \ln\frac{\mu_f^2}{\mu^2}+8
\right)\Bigg)\Bigg],
\nonumber 
\end{eqnarray}
where $b_0=-4\pi\beta_0=11-2 n_f/3$ and 
$b_1=-(4\pi)^2\beta_1=102-38 n_f/3$. Note that the logarithms 
disappear when the coupling is normalised at the scale $\mu_f$, 
which follows from the fact that the potential is physical and 
independent of $\mu$. At $\mu_f=1$-$1.5\,$GeV, a typical scale relevant 
for bottom quarks, the third-order term already exceeds the 
second-order term. This is not a point of concern as the 
series expansion of $\delta m(\mu_f)$ is expected to behave badly 
and we are interested only in the combinations $V(r)+2\delta m(\mu_f)$ 
and $m_{\rm pole}-\delta m(\mu_f)$, both of which have better 
behaved series expansions. The subtracted Coulomb potential 
$V(r)+2\delta m(\mu_f)$ at leading order in 
$\alpha_s$ is given by 
\begin{equation}
V(r)=-\frac{C_F\alpha_s(\mu_r)}{r}\left\{1-\frac{2}{\pi}\,
\frac{\alpha_s(\mu_f)}{\alpha_s(\mu_r)}\,\mu_f r\right\} 
\end{equation}
with $\mu_r=e^{-\gamma_E}/r$. 
To see the numerical effect of the subtraction, we choose the 
values $r=1/(20\,\mbox{GeV})$ and $\mu_f=3\,$GeV ($n_f=5$) as would 
be appropriate to $t\bar{t}$ production and compare the 
subtracted and unsubtracted Coulomb potential. The result, 
including the known higher-order corrections, is 
\begin{equation}
\label{vsu}
V(r,\mu_f) = -\frac{C_F\alpha_s(\mu_r)}{r}\left\{
0.86 + 0.16\,\frac{\alpha_s(\mu_r)}{\pi} + 
13.64\,\left(\frac{\alpha_s(\mu_r)}{\pi}\right)^2+\ldots
\right\},
\end{equation}
as compared to
\begin{equation}
V(r) = -\frac{C_F\alpha_s(\mu_r)}{r}\left\{
1+ 1.19\,\frac{\alpha_s(\mu_r)}{\pi} + 
32.93\,\left(\frac{\alpha_s(\mu_r)}{\pi}\right)^2+\ldots
\right\}.
\end{equation}
The convergence of the series is improved and the strength of 
the potential is reduced. For bottom systems one observes a 
similar effect, although the requirement that 
$\mu_f > \Lambda_{\rm QCD}$ does not allow us to choose 
$\mu_f r$ as small as we would like. 

\begin{table}[b]
\addtolength{\arraycolsep}{0.2cm}
\renewcommand{\arraystretch}{1.2}
$$
\begin{array}{c|c|c}
\hline\hline
\mu_f/M & r_1^{\rm PS} & r_2^{\rm PS} \\ 
\hline
0    & 1.33  & 13.44-1.04 n_f \\
1/25 & 1.28  & 12.07-0.95 n_f \\
1/5  & 1.07  & 8.927-0.73 n_f \\
1/4  & 1     & 8.207-0.68 n_f \\
1/3  & 0.89  & 7.165-0.61 n_f \\
\hline\hline
\end{array}
$$
\caption{\label{tab1} \small First and second-order (in $\alpha_s(M)/\pi$) 
coefficients in the relation between the PS mass and the 
$\overline{\rm MS}$ mass (\ref{pstoms}). For $\mu_f=0$ the 
PS mass coincides with the pole mass.}
\end{table}
Since the relation of the pole mass to the $\overline{\rm MS}$ mass 
is known only to second order \cite{GBGS90}, we can only make use 
of $\delta m(\mu_f)$ to second order to express the potential-subtracted 
mass $m_{\rm PS}(\mu_f)=m_{\rm pole}-\delta m(\mu_f)$ in terms 
of $M\equiv m_{\overline{\rm MS}}( m_{\overline{\rm MS}})$. 
The result is 
\begin{eqnarray}
\label{pstoms}
m_{\rm PS}(\mu_f) &=& M\,\Bigg\{1+\frac{4\alpha_s(M)}{3\pi} 
\left[1-\frac{\mu_f}{M}\right] 
\nonumber\\
&&\hspace*{-1cm}+ \left(\frac{\alpha_s(M)}{\pi}
\right)^2\left[K-\frac{\mu_f}{3 M}\left(a_1-b_0\left[
\ln\frac{\mu_f^2}{M^2}-2\right]\right)\right]+\ldots\Bigg\},
\end{eqnarray}
where $K=13.44-1.04 n_f$ from\footnote{Note that 
$m_{\overline{\rm MS}}(m_{\rm pole})$ is used in \cite{GBGS90} 
and the different normalisation point for the quark mass affects 
the second order coefficient in the relation to $m_{\rm pole}$.} 
\cite{GBGS90} and $a_1=10.33-1.11 n_f$.
Numerical values for the first and second-order coefficients 
for various values of $\mu_f/M$ are given in Table~\ref{tab1}. 
For small values of $\mu_f/M$ as relevant to $t\bar{t}$ production 
the series is not very different from the series for the 
pole mass, reflecting the fact that at this order 
both series still converge well. 
In absolute terms the difference between the pole and PS mass 
may still amount to several hundred MeV, which is significant 
close to threshold. In fact the effect of 
the subtraction is far from small on the potential even for $t\bar{t}$ 
production as illustrated by (\ref{vsu}) above, because 
$\mu_f$ has to be compared to the scale $m_t v/2\sim m_t \alpha_s/2$ 
in case of the potential. For ratios of $\mu_f/M$ that may 
be contemplated for bottom quarks near threshold the second-order 
coefficient in (\ref{pstoms}) is already considerably smaller 
than the one in the relation of $m_{\rm pole}$ to $M$ and the 
convergence of the series expansion is improved. This lends support 
to the hypothesis that the IR cancellations between 
$m_{\rm pole}$ and $\delta m(\mu_f)$ occur not only asymptotically 
but already at second-order. 
Eq.~(\ref{vsu}) and (\ref{pstoms}) taken together suggest that 
threshold calculations at NNLO formulated in terms of the subtracted 
potential and the PS mass exhibit reduced NNLO corrections and 
that the PS mass can indeed 
be reliably related to the $\overline{\rm MS}$ 
mass.\\
 
{\bf \noindent 6. Conclusion}\\[-0.2cm]

\noindent
In this paper we proposed that perturbative calculations of heavy 
quark properties near threshold should not use the pole mass 
but a subtracted mass together with a subtracted potential. This 
eliminates one source of large corrections in perturbation theory, 
related to small momentum contributions, 
although we cannot exclude large corrections due to other 
reasons. The numerical estimates presented above 
suggest, however, that the convergence is indeed improved. We emphasise 
that the potential-subtracted mass is gauge-invariant, because 
it is derived from the pole mass and an integral over the momentum 
space potential, both of which are gauge-invariant.

The crucial point is that heavy quark cross sections near threshold 
are in fact less sensitive to long distances than the quark 
pole mass parameter. This follows from the observation that the Coulomb 
potential in momentum space is less sensitive to long distances 
than the potential in coordinate space and that the large 
perturbative corrections to the potential in coordinate space 
cancel to all orders in perturbation theory with those to the 
pole mass, because of an exact cancellation of the small 
momentum behaviour of the respective Feynman integrals. That the pole mass
is not relevant for physical quantities involving top quarks 
is quite obvious, because the width of order $1.5\,$GeV provides 
an intrinsic cut-off for long-distance effects \cite{BDKKZ86}. 
In particular, the 
location of the resonance-like peak in the production cross section 
is not a direct measure of the top quark pole mass despite the fact 
that top quarks do not hadronise. It is however less obvious 
that the pole mass is not even relevant for (quasi-) stable quarks 
near threshold.

Making use of the results of \cite{Pet97} we derived the 
mass subtraction term to order $\alpha_s^3$ and the relation 
between the potential-subtracted (PS) mass and the 
$\overline{\rm MS}$ mass to order $\alpha_s^2$. These 
relations provide the link with other physical quantities 
involving heavy quarks. In particular, they allow us to 
determine directly the bottom and top $\overline{\rm MS}$ masses 
from bottomonium sum rules and the $t \bar{t}$ cross section, 
thus obviating large NNLO corrections that appear when these 
quantities are expressed in terms of pole masses 
\cite{HT98,MY98,PP98,Hoa98}. One may hope that this leads 
to more accurate quark mass determinations than for the pole masses, 
whose accuracy is limited to order $\Lambda_{\rm QCD}$  
by long-distance effects \cite{BB94,BSUV94} independent of 
the process utilised to determine them. We will report on these 
applications in a forthcoming publication. 

One cannot use the $\overline{\rm MS}$ masses themselves for 
threshold problems, because they differ from the pole masses by 
an amount of order $m\alpha_s$. This causes singular terms 
of order $(\alpha_s/v^2)^k$ to appear in perturbative expansions. 
The all-order resummation of these terms 
leads one back to the pole mass. It is necessary to introduce 
a factorisation scale $\mu_f$ and to choose a mass definition (the 
PS mass) that differs from the pole mass by an amount smaller 
than the typical energies of a Coulomb system, while at the 
same time not being too sensitive to confinement effects. 
This leads to a linear dependence on the subtraction scale 
$\mu_f$. The use of a heavy quark mass with a linear factorisation 
scale dependence has been repeatedly advocated by 
Bigi {\em et al.} (see \cite{BSUV94} and the review 
\cite{BSU97}). In \cite{CMU97} a 
mass subtraction term $\bar{\Lambda}(\mu_f)$ (the analogue 
of our $\delta m(\mu_f)$) is derived to order $\alpha_s^2$ from 
certain integrals over the spectral densities of heavy-light 
quark currents. This mass subtraction differs from 
(\ref{deltam}) already at order $\alpha_s$. This does not 
imply an inconsistency, since the necessary requirement is 
only that the long-distance sensitive regions cancel 
asymptotically in large orders. On the other hand, it seems 
to us that a subtraction based on the heavy quark Coulomb potential 
is most natural (and technically simplest) not only for 
threshold problems involving two heavy quarks, since, as 
observed in \cite{BSUV94}, the leading 
long-distance sensitive contribution 
to the pole mass is in fact conceptually related to the 
Coulomb interaction.\\

\noindent
{\em Note added:} After this paper was completed, we 
received Ref.~\cite{HSSW98}, which addresses related questions. 
The authors also note that linear 
IR sensitivity cancels in $2m_{\rm pole}+V(r)$ and demonstrate 
this at the 1-loop order.\\

\noindent
{\em Acknowledgements.} I thank G.~Buchalla, A.~Signer and V.A.~Smirnov 
for comments on the manuscript and N.~Uraltsev for correspondence.


\end{document}